\documentclass[prl,aps,twocolumn,preprintnumbers,showpacs]{revtex4}

\begin{document}
\title{Kraus representation for density operator of arbitrary open qubit system}
\author{D. M. Tong $^1$, Jing-Ling Chen $^1$, L. C. Kwek$^{1,2}$ and C. H. Oh$^1$
}
\address{$^1$Department of Physics, National University of
Singapore, 10 Kent Ridge Crescent, Singapore 119260 \\
$^2$ National Institute of Education, Nanyang Technological
University, 1 Nanyang Walk, Singapore 639798 }
\date{\today}

\begin{abstract}
We show that the time evolution of density operator of open qubit
system can always be described in terms of the Kraus
representation. A general scheme on how to construct the Kraus
operators for an open qubit system is proposed, which can be
generalized to open higher dimensional quantum systems.
\end{abstract}
\pacs{03.65.Yz, 03.65.Ca} \maketitle

\section{Introduction}
It is well-known that for a closed quantum system, its time
evolution can be described by a unitary operator. However, for an
open system, the time evolution is not necessarily unitary. The
evolution of an open system is usually described by the Kraus
representation \cite{Kraus}. Since a real physical system is
generally entangled with its environment, the proper
understanding on the nature of the Kraus representation for an
open system is important and
useful\cite{Kraus,Preskill,Pechukas,Pomero,Bouda,
Philip,Gen,Peter, Salgado, Hayashi}, especially in quantum
information processing.

The Kraus representation of an open system is usually constructed
by considering a larger closed system denoted as $S_{ie}$,
comprising of the interested system $S_i$ and its environment
$S_e$. Let $\rho_{ie}(t)$, $\rho_i(t)$ and $\rho_e(t)$ be the
density matrices of $S_{ie}$, $S_i$ and $S_e$ respectively, where
$\rho_i(t)={\rm tr}_e[\rho_{ie}(t)]$ and $\rho_e(t)={\rm
tr}_i[\rho_{ie}(t)]$, and $\rho_{ie}(0)$, $\rho_i(0)$ and
$\rho_e(0)$ represent the corresponding initial states
respectively at $t=0$. As the combined system is a closed one,
its evolution is unitary,
\begin{eqnarray}
\rho_{ie}(t)=U_{ie}(t)\rho_{ie}(0)U_{ie}(t)^+, \label{rhoiet}
\end{eqnarray}
where $U_{ie}(t)$ is the unitary operator. The interested system,
as an open one, then evolves in the following way
\begin{eqnarray}
\rho_i(t)={\rm tr}_e\{U_{ie}(t)\rho_{ie}(0)U_{ie}(t)^+\}.
\label{rhoit}
\end{eqnarray}
 If the above equation can be equivalently expressed in the form
\begin{eqnarray}
\rho_i(t)=\sum\limits_{\mu \nu} M_{\mu \nu}(t)\rho_i(0)M_{\mu
\nu}(t)^+, \label{rhoitm}
\end{eqnarray}
where $M_{\mu\nu}(t)$ satisfy
\begin{eqnarray}
\sum\limits_{\mu \nu} M_{\mu \nu}(t)^+M_{\mu \nu}(t)=I,
\label{mm1}
\end{eqnarray}
it is said that the evolution of $\rho_i(t)$ has the form of the
Kraus representation.

It is obvious that $\rho_i(t)$ always has the Kraus
representation for arbitrary $U_{ie}(t)$ if $\rho_{ie}(0)$ is
{\it{factorable}} \cite{Bouda}, {\it {i.e.}}
$\rho_{ie}(0)=\rho_i(0)\otimes \rho_e(0)$, which means that there
is no initial correlation between the open system and its
environment. To show this, we can take $\rho_e(0) =\sum_\nu
\sqrt{p_\nu} |\nu_e\rangle\langle \nu_e |$ and let
\begin{eqnarray}
 M_{\mu\nu}(t)=\langle \mu_e |\sqrt{p_\nu} U_{ie}(t)|\nu_e \rangle,
\label{ms}
\end{eqnarray}
where  $|\mu_e \rangle $,  $|\nu_e \rangle $ ( $ \mu,~\nu=0, 1,
... ,k-1$) are the orthonormal bases of $S_e$, and $k$ is the
dimension of $S_e$, one will find that $M_{\mu\nu}(t)$ defined by
Eq. (\ref{ms}) satisfy Eqs. (\ref{rhoitm}) and (\ref{mm1}).

The issue is that whether $\rho_i(t)$ still has the form of the
Kraus representation when $\rho_{ie}(0)$ is not
{\it{factorable}}, which means that the initial correlations
between $S_i$ and $S_e$ are present. Or in other words, can one
always find the Kraus representation of an open system for
arbitrary initial state $\rho_{ie}(0)$ and arbitrary unitary
operator $U_{ie}(t)$? Recently, some papers\cite{Peter, Salgado,
Hayashi} have contributed to the issue. \u Stelmachovi\u c
{\it{et al.}} \cite{Peter} investigated the role of the initial
correlations between the open system and its environment and
showed that a map based on the reduced dynamics in the presence
of initial correlations can't be described by the form of the
Kraus representation because an additional inhomogeneous part
appears. Salgado {\it{et al.}} \cite{Salgado} pointed out that
$\rho_i(t)$ still has the Kraus representation even in the
presence of any initial correlation if the evolution is local,
namely $U_{ie}(t)=U_i(t)\otimes U_e(t)$. In a very recent paper
\cite{Hayashi}, Hayashi {\it et al.} examined the validity of the
Kraus representation in the presence of initial correlations and
concluded that the dynamical map for an open system reduced from
a combined system with an arbitrary initial correlation takes the
form of the Kraus representation if and only if the joint
dynamics is locally unitary.

To arrive at the above conclusion, an operator $\rho_{cor}(0)$,
called the correlation operator, was introduced through the
definition $\rho_{cor}(0)\equiv \rho_{ie}(0)-\rho_i(0)\otimes
\rho_e(0)$. Equation (\ref{rhoit}) can then be recast to the
following form
\begin{eqnarray}
\rho_i(t)&=&{\rm tr}_e\{U_{ie}(t)\rho_i(0)\otimes
\rho_e(0)U_{ie}(t)^+\} \nonumber \\&&
+{\rm tr}_e\{U_{ie}(t)\rho_{cor}(0)U_{ie}(t)^+\}\nonumber \\
&=&\sum\limits_\mu M_{\mu \nu}(t)\rho_i(0)M_{\mu \nu}(t)^+ +
\delta \rho_i(t), \label{rhodelta}
\end{eqnarray}
where
\begin{eqnarray}
\delta \rho_i(t)={\rm tr}_e\{U_{ie}(t)\rho_{cor}(0)U_{ie}(t)^+\}.
\label{delta}
\end{eqnarray}
The analysis in Refs. \cite{Peter, Salgado, Hayashi} is based on
the idea that $\rho_i(t)$ has the Kraus representation if and
only if $\delta \rho_i(t)=0$. Clearly, $\rho_i(t)$ has the form
of Eq. (\ref{rhoitm}) if $\delta \rho_i(t)=0$ and the Kraus
operators are given by Eq. (\ref{ms}). However, noticing that
Kraus operators are highly nonunique, one may start wondering
whether $\rho_i(t)$ has an alternative form of the Kraus
representation even if $\delta \rho_i(t)\neq 0$, because there
may exist Kraus operators $\tilde {M}_{\mu \nu}(t)$ such that
\begin{eqnarray}
\rho_i(t)&=&\sum\limits_\mu M_{\mu \nu}(t)\rho_i(0){M_{\mu
\nu}(t)}^+ + \delta \rho_i(t)\nonumber\\&=&\sum\limits_\mu
\tilde{M}_{\mu\nu}(t)\rho_i(0){\tilde{M}_{\mu\nu}}(t)^+,
\end{eqnarray}
and $\sum_{\mu \nu} {\tilde{M}_{\mu\nu}}(t)^+
\tilde{M}_{\mu\nu}(t)=I$. $\tilde{M}_{\mu\nu}$ may not be
calculated from Eq. (\ref{ms}), but they need to have the
properties of Kraus operators, which ensure the map defined by
them to be hermitian, trace preserving and positive.

We consider this problem in the present paper. Our investigation
focuses on the open qubit system. The paper is organized as
follows. In Sec. II, an example is provided to show that the
alternative Kraus representation really exists even if $\delta
\rho_i(t)\neq 0$. In Sec. III, we propose a general approach on
how to construct Kraus operators for an arbitrary open qubit
system. We end with some discussions in the final section.

\section{Kraus representation with nonzero $\delta \rho_i(t)$}
In this section, by providing an example, we show that
$\rho_i(t)$ may still have an alternative  form of the Kraus
representation even if $\delta \rho_i(t)\neq 0$. We choose the
same model as that Ref.\cite{Hayashi} has used. That is, we
consider a combined system composed of two spin-1/2 subsystems
with the interaction Hamiltonian $H_{ie}=\sigma_x\otimes
\frac{1}{2}({\bf 1}-\sigma_z)+ {\bf 1} \otimes
\frac{1}{2}(1+\sigma_z)$~, where $\sigma_x$ and $\sigma_z$ are
Pauli spin operators. In this model, the first qubit plays the
role of the open system while the second qubit plays the role of
the environment. The  interaction described by the Hamiltonian
corresponds to the well-known controlled-NOT
gate\cite{Preskill,Peter}. The unitary evolution operator is
given by $U_{ie}(t)=e^{-iH_{ie}t}$, explicitly
\begin{eqnarray}
U_{ie}(t)=\left(\begin{array}{cccc}e^{-it}&0&0&0\\0&\cos
t&0&-i\sin t\\0&0&e^{-it}&0\\0&-i\sin t&0&\cos
t\end{array}\right). \label{Uie}
\end{eqnarray}
In the model considered, $\rho_{ie}$ is a $4\times 4$ matrix
while  $\rho_i$ and $\rho_e$ are $2\times 2$ matrices. For
simplicity, we take the initial state of the combined system as
\begin{eqnarray}
\rho_{ie}(0)=\left(\begin{array}{cccc}
\frac{1-r_0}{2}&0&0&0\\0&0&0&0\\0&0&0&0\\0&0&0&\frac{1+r_0}{2}\end{array}\right),
\label{rhoie0}
\end{eqnarray}
where $r_0\in (0,~ 1)$ is a real parameter. Noting that at
$r_0=0$ or $1$, $\rho_{ie}(0)$ is {\it{factorable}}, and the
Kraus representation certainly exists, we needn't consider these
two cases.

It is easy to obtain the initial reduced density matrices of
$S_i$ and $S_e$ as
\begin{eqnarray}
&\rho_i(0)&=tr_e\rho_{ie}(0)=
\frac{1}{2}({\bf 1}-r_0 \sigma_z),\nonumber \\
&\rho_e(0)&=tr_i\rho_{ie}(0)=\frac{1}{2}({\bf 1} -r_0 \sigma_z),
\label{rhorho0}
\end{eqnarray}
and the correlation operator is
\begin{eqnarray}
\rho_{cor}(0)=\frac{1}{4}(1-r_0^2) \; \sigma_z \otimes \sigma_z.
\label{rhocor2}
\end{eqnarray}
From Eqs. (\ref{rhoit}), (\ref{Uie}) and (\ref{rhoie0}), we get
the density matrix of the system $S_i$,
\begin{eqnarray}
\rho_i(t)=\frac{1}{2}\left(\begin{array}{cc}1+\sin^2t-r_0\cos^2t&~-i(1+r_0)\sin t\cos t\\i(1+r_0)\sin t\cos t&(1+r_0)\cos^2 t\end{array}\right).\nonumber\\
\label{rhoit2}
\end{eqnarray}
Substituting Eqs. (\ref{Uie}) and (\ref{rhocor2}) into Eq.
(\ref{delta}), one gets
\begin{eqnarray}
\delta
\rho_i(t)=\frac{1}{4}(1-r_0^2)\left(\begin{array}{cc}2\sin^2
t&~-i\sin 2t\\i\sin 2t&~-2\sin^2 t\end{array}\right).
\label{rhodelta2}
\end{eqnarray}
We see that $\delta \rho_i(t)$ is, in general, non-zero. However,
the Kraus representation of $\rho_i(t)$ still exists. One can
verify that the following expressions hold,
\begin{eqnarray}
\rho_i(t)=\sum\limits_{\mu=0}^1 M_\mu (t) \rho_i(0)M_\mu (t)^+,
\label{rhorho}
\end{eqnarray}
\begin{eqnarray}
\sum\limits_{\mu=0}^1 M_\mu (t)^+M_\mu (t)=I, \label{I}
\end{eqnarray}
with
\begin{widetext}
\begin{eqnarray}
&M_0 (t)&=\frac{1}{\sqrt{2r_t(1+r_0)}}\left(\begin{array}{cc}-\sqrt{(1+r_0)(r_t+\sin^2t-r_0\cos^2t)}&~~i\sqrt{(1-r_t)(r_t-\sin^2t+r_0\cos^2t)}\\-i\sqrt{(1+r_0)(r_t-\sin^2t+r_0\cos^2t)}& ~~\sqrt{(1-r_t)(r_t+\sin^2t-r_0\cos^2t)}\end{array}\right),\nonumber \\
&M_1
(t)&=\frac{\sqrt{r_t+r_0}}{\sqrt{2r_t(1+r_0)}}\left(\begin{array}{cc}0&~~\sqrt{r_t+\sin^2t-r_0\cos^2t}\\0&~~i\sqrt{r_t-\sin^2t+r_0\cos^2t}\end{array}\right),
\label{MM}
\end{eqnarray}\end{widetext}
where $r_t=\sqrt{\sin^2t+r_0^2\cos^2t}$, $M_0 (t)$ and $M_1 (t)$
are the Kraus operators.

The map defined by Eq. (\ref{MM}) ensures $\rho_i(t)$ hermitian,
trace preserving and positive. The evolution of the system $S_i$
obeys Eq. (\ref{rhorho}) while the combined system evolves under
unitary operator $U_{ie}(t)$ given by Eq. (\ref{Uie}). This
example has showed that even if $\delta \rho_i(t)\neq 0$,
$\rho_i(t)$ can still be written as the form of Kraus
representation.

\section{Kraus representation for arbitrary density operator}
From Eq. (\ref{rhodelta}), we see that the state $\rho_i(t)$
cannot be written in the form of the Kraus representation with
the Kraus operators defined by Eq. (\ref{ms}) if  $\delta
\rho_i(t)\neq 0$. However, the example in section II illustrates
that there may exist an alternative form of the Kraus
representation even if  $\delta \rho_i(t)\neq 0$.  This
encourages us to conjecture that the time evolution of the
density operator can always have the Kraus representation
irrespective of the forms of initial state and evolution path. In
this section, we will prove that it is true that $\rho_i(t)$
always can be connected with its initial state $\rho_i(0)$ by
Kraus operators.

Let us begin by considering an arbitrary evolution of an open
qubit system with arbitrary initial state. The most general
initial state for an open qubit system can be written as
\begin{eqnarray}
\rho_i(0)=\frac{1}{2}(1+{\bf{r_0 \cdot \sigma}})=\frac{1}{2}
\left(\begin{array}{cc}
1+r_0\cos\theta_0&~r_0\sin\theta_0~e^{-i\phi_0}
\\r_0\sin\theta_0~e^{i\phi_0}&~1-r_0\cos\theta_0
\end{array}\right),
\label{rho03}
\end{eqnarray}
and the most general evolution of the system reads as
\begin{eqnarray}
\rho_i(t)=\frac{1}{2}(1+{\bf{r \cdot \sigma}})=\frac{1}{2}
\left(\begin{array}{cc} 1+r\cos\theta&~r\sin\theta~e^{-i\phi}
\\r\sin\theta~e^{i\phi}&~1-r\cos\theta
\end{array}\right),
\label{rhot3}
\end{eqnarray}
where ~$r=r(t),$~$\theta=\theta (t) ,\phi =\phi (t), $~depending
on time $t$, and $r(0)=r_0,~\theta (0)= \theta_0 ,~\phi(0) =
\phi_0$. $0\leq r \leq 1,~0\leq \theta \leq \pi,~0\leq \phi \leq
2\pi.$ We want to  show that there always exist the Kraus
operators $M_\mu (t) $, such that
\begin{eqnarray}
\rho_i(t)=\sum\limits_\mu M_\mu (t)\rho_i(0){M_\mu (t)} ^+,
\label{rhotm3}
\end{eqnarray}
\begin{eqnarray}
\sum\limits_\mu M_\mu (t)^+M_\mu (t)=I, \label{mm3}
\end{eqnarray}
where we have used $M_{\mu}(t)$, instead of $M_{\mu\nu}(t)$, to
denote the Kraus operators. To find the Kraus operators,  one may
write $M_\mu (t)$ as $2\times 2$ matrices with undetermined
elements and one may then directly solves  Eqs. (\ref{rhotm3})
and (\ref{mm3}) to determine the matrices. However, it is too
difficult to do in that way. As diagonal matrix is, in general,
easier to handle than non-diagonal ones, we first diagonalize the
density matrices $\rho_i(0)$ and $\rho_i(t)$ by unitary
transformations,
\begin{eqnarray}
\rho_i(0)=U_1 \rho'_i(0)U_1^+, ~~~~
\rho_i(t)=U_2(t)\rho'_i(t)U_2(t)^+. \label{rhouu}
\end{eqnarray}
The eigenvalues of $\rho_i(0)$ and $\rho_i(t)$ make up the
entries of the diagonalized matrices $\rho'_i(0)$ and
$\rho'_i(t)$ respectively. And their orthogonal vectors make up
the columns of the unitary matrices $U_1$ and $U_2$ respectively.
In this way, the diagonalized matrices can be written as
\begin{eqnarray}
\rho'_i(0)=\frac{1}{2}(1+{\bf{r'_0 \cdot \sigma}})=\frac{1}{2}
\left(\begin{array}{cc} 1-r_0&0
\\0&1+r_0
\end{array}\right),
\label{r0}
\end{eqnarray}
\begin{eqnarray}
\rho'_i(t)=\frac{1}{2}(1+{\bf{r' \cdot
\sigma}})=\frac{1}{2}\left(\begin{array}{cc} 1+r&0
\\0&1-r
\end{array}\right),
\label{rt}
\end{eqnarray}
where ${\bf{r'_0}}$ and ${\bf{r'}}$ are defined as
${\bf{r'_0}}=(0,~0,~-r_0)$ and ${\bf{r'}}=(0,~0,~r)$
respectively, and the corresponding unitary transformation
matrices are
\begin{eqnarray}
U_1=\left(\begin{array}{cc}
-\sin\frac{\theta_0}{2}&~\cos\frac{\theta_0}{2}e^{-i\phi_0}
\\\cos\frac{\theta_0}{2}e^{i\phi_0}&~\sin\frac{\theta_0}{2}
\end{array}\right),
\label{ut0}
\end{eqnarray}
\begin{eqnarray}
U_2=\left(\begin{array}{cc}
\cos\frac{\theta}{2}&~-\sin\frac{\theta}{2}e^{-i\phi}
\\ \sin\frac{\theta}{2}e^{i\phi}&~\cos\frac{\theta}{2}
\end{array}\right).
\label{utt}
\end{eqnarray}
If we can find such operators $M'_\mu (t)$ that satisfy $
\rho_i'(t)=\sum_\mu M'_\mu(t) \rho_i'(0){M'_\mu}(t)^+$ and
$\sum_\mu {M'_\mu}(t)^+M'_\mu(t)=I$, then, the Kraus
representation of $\rho_i(t)$ can be realized by letting
\begin{eqnarray}
M_\mu (t)=U_2M'_\mu(t) U_1^+. \label{mm}
\end{eqnarray}
Since $\rho'_i(t)$ and $\rho'_i(0)$ are diagonal, the operators
$M'_\mu(t)$  are easy to find. There are infinite choices of this
kind of Kraus operators. Without loss of generality, we may choose
them as
\begin{eqnarray}
&M'_0 (t)&= \left(\begin{array}{cc} 1&0
\\0&\sqrt{\frac{1-r}{1+r_0}}
\end{array}\right),\nonumber \\
&M'_1 (t)&= \left(\begin{array}{cc} 0&\sqrt{\frac{r+r_0}{1+r_0}}
\\0&0
\end{array}\right).
\label{mm01}
\end{eqnarray}
Substituting Eqs. (\ref{ut0}), (\ref{utt}) and (\ref{mm01}) into
Eq. (\ref{mm}), we obtain the Kraus operators $ M_\mu (t)
$,\begin{widetext}
\begin{eqnarray}
&M_0 (t)&= \left(\begin{array}{cc}
-\cos\frac{\theta}{2}\sin\frac{\theta_0}{2}-\sqrt{\frac{1-r}{1+r_0}}\sin\frac{\theta}{2}\cos\frac{\theta_0}{2}e^{i(\phi_0-\phi)}&~\cos\frac{\theta}{2}\cos\frac{\theta_0}{2}e^{-i\phi_0}-\sqrt{\frac{1-r}{1+r_0}}\sin\frac{\theta}{2}\sin\frac{\theta_0}{2}e^{-i\phi}
\\-\sin\frac{\theta}{2}\sin\frac{\theta_0}{2}e^{i\phi}+\sqrt{\frac{1-r}{1+r_0}}\cos\frac{\theta}{2}\cos\frac{\theta_0}{2}e^{i\phi_0}&~\sin\frac{\theta}{2}\cos\frac{\theta_0}{2}e^{i(\phi-\phi_0)}+\sqrt{\frac{1-r}{1+r_0}}\cos\frac{\theta}{2}\sin\frac{\theta_0}{2}
\end{array}\right),\nonumber \\
&M_1 (t)&= \sqrt{\frac{r+r_0}{1+r_0}}\left(\begin{array}{cc}
\cos\frac{\theta}{2}\cos\frac{\theta_0}{2}e^{i\phi_0}&~\cos\frac{\theta}{2}\sin\frac{\theta_0}{2}\\\sin\frac{\theta}{2}\cos\frac{\theta_0}{2}e^{i(\phi+\phi_0)}&~\sin\frac{\theta}{2}\sin\frac{\theta_0}{2}e^{i\phi}
\end{array}\right).
\label{mm02}
\end{eqnarray}\end{widetext}
$M_0 (t)$ and $M_1 (t)$ satisfy Eqs. (\ref{rhotm3}) and
(\ref{mm3}). The Kraus representation given by $M_0 (t)$ and $M_1
(t)$ in expression (\ref{mm02}) does ensure $\rho_i(t)$ be
hermitian, trace preserving and positive. So, no matter what the
forms of $U_{ie}(t)$ and $\rho_{ie}(0)$ are, there always exist
the Kraus operators connecting $\rho_i(t)$ with $\rho_i(0)$. For
any given $\rho_{ie}(0)$ and $U_{ie}(t)$, the Kraus operators
$M_\mu (t)$ can be calculated by diagonalizing the reduced
matrices ${\rm tr}_e\{U_{ie}(t)\rho_{ie}(0)U_{ie}(t)^+\}$ and
${\rm tr}_e\rho_{ie}(0)$. One general expression of the Kraus
operators is given by Eq. (\ref{mm02}), with which the Kraus
representation of $\rho_i(t)$ is obtained by Eq. (\ref{rhotm3}).

So far, we have proved that the time evolution of an density
operator of open qubit system always has the Kraus
representation. At the same time, we have put forward a general
approach for constructing the Kraus operators for arbitrary
evolution. From physical point of view, the above process of
finding the Kraus operators means that we first align the Bloch
vectors ${\bf{r}}$ and ${\bf{r_0}}$ in Bloch sphere along the $z$
axis by using $U_1$ and $U_2$ respectively, find the Kraus
representation of ${\bf{r'}}$ with $\bf{r'_0}$, and then revers
${\bf{r'}}$ and $\bf{r'_0}$ back to ${\bf{r}}$ and $\bf{r_0}$, to
obtain the Kraus representation of ${\bf{r}}$ with $\bf{r}_0$.
The model in sec.II is just an example of applying this approach
to solve the Kraus representation. In fact, expression (\ref{MM})
is calculated in this way.

\section{Discussions}
We have shown that the time evolution of the density operator of
an open qubit system can always have the Kraus representation. A
scheme on how to construct the Kraus representation is proposed.
One general expression of the Kraus representation for an
arbitrary evolution is provided by Eqs. (\ref{rhotm3}),
(\ref{mm3}) and (\ref{mm02}). Since the expressions of the Kraus
operators are not unique, the form given by Eq. (\ref{mm02}) is
only one kind of them.  The equivalent expressions of the Kraus
operators can be written down as $\tilde{M}_\mu(t)=\sum_\nu
M_\nu(t) V_{\mu\nu}$, where $V_{\mu\nu}$ are the elements of an
arbitrary unitary matrix.

Refs. \cite{Peter, Salgado, Hayashi} have investigated the
possibility of the Kraus representation for an open system with
initial correlations between the system and its environment and
some important conclusions have been derived. As a supplement,
the present paper studies the existence of an operator-sum
representation for an arbitrary given evolution of density
operator. Our result shows that an arbitrary evolution of the
state can always be written as the form of the Kraus
representation. The Kraus operators can be calculated by Eq.
(\ref{ms}) if $\delta \rho_i(t) = 0$. However, they cannot be
expressed explicitly in the form of Eq. (\ref{ms}) if $\delta
\rho_i(t)\neq 0$. For this latter case, they can still be obtained
by the approach described in the current paper. Moreover, $M_\mu
(t)$ are generally dependent on the initial state and there does
not exist a universal form of Kraus operators for all different
initial states. 

This approach can be generalized to higher dimensional quantum
systems. The procedure for higher dimensional systems is similar
to the qubit case but may be more complicated. In fact, the
density matrix $\rho_i(t)$ with the parameter $t$ and $\rho_i(0)$
can always be diagonalized by unitary transformations $U_1$ and
$U_2$ respectively, regardless of the dimensions of the density
matrices. It is easy to find the Kraus operators $M'_\mu (t)$ of
diagonal density matrix $\rho'_i(t)$ with diagonal initial density
matrix $\rho'_i(0)$ though it is difficult to find the Kraus
representation of an arbitrary density matrix with arbitrary
initial conditions. By Eq. (\ref{mm}), using $U_1$, $U_2$ and
$M'_\mu (t)$, the Kraus representation of $\rho_i(t)$ is obtained.
Certainly, as the dimensions of the density matrices become
larger, solving for the Kraus operator may become more formidable.

\section*{Acknowledgments}
The work was supported by NUS Research Grant No.
R-144-000-071-305. JLC acknowledges financial support from
Singapore Millennium Foundation.

\end{document}